\documentclass[pra,aps,twocolumn,superscriptaddress,showpacs]{revtex4-2}
\usepackage{amsmath,amssymb,graphicx}
\usepackage{float}
\usepackage{amsfonts}
\usepackage{amsmath}
\usepackage{amssymb}
\usepackage{dsfont}
\usepackage{bbold}
\usepackage{bm}

\usepackage{graphicx}% Include figure files
\usepackage{dcolumn}% Align table columns on decimal point
\usepackage[colorlinks=true,linkcolor=blue,citecolor=blue,urlcolor=blue]{hyperref}
\usepackage{xcolor}
\usepackage{braket}
\usepackage{textcomp}

\begin{document}
\title{Emerging Non-Hermitian Topology in a Chiral Driven-Dissipative Bose-Hubbard Model}
\author{Laszlo Rassaert}
\email{laszlo.rassaert@ens-lyon.fr}
\affiliation{Laboratory of Physics, University of Lyon, Ens de Lyon, CNRS, Lyon, France}
\author{Tom\'as Ramos}
\affiliation{Instituto de F\'{\i}sica Fundamental IFF-CSIC, Calle Serrano 113b, 28006 Madrid, Spain.}
\author{Tommaso Roscilde}
\affiliation{Laboratory of Physics, University of Lyon, Ens de Lyon, CNRS, Lyon, France}
\author{Diego Porras}
\email{diego.porras@csic.es}
\affiliation{Instituto de F\'{\i}sica Fundamental IFF-CSIC, Calle Serrano 113b, 28006 Madrid, Spain.}

\date{\today}

\begin{abstract}
We introduce a driven-dissipative Bose-Hubbard chain describing coupled lossy photonic modes, in which time-reversal symmetry is broken by a coherent drive with a uniform phase gradient. 
We investigate this model by means of a Gaussian variational ansatz and numerically prove that the steady-state solution is stabilized by an inhomogeneous profile of the driving amplitude, which damps out boundary effects. 
Our calculations unveil a non-equilibrium phase diagram showing low- and high-density phases for photons separated by a phase coexistence region in which the system exhibits the phenomenon of topological amplification 
and is characterized by a finite non-Hermitian winding number. 
Our work shows the emergence of non-Hermitian topological phases in an interacting model that can be naturally implemented with superconducting circuits.
\end{abstract}

\maketitle

\emph{Introduction.-}
The study of topology in condensed matter physics has unveiled a plethora of exciting phenomena like the quantum Hall effect \cite{klitzing80prl} and topological insulators with time-reversal symmetry \cite{hasan2010rmp,qi2011rmp}.
Those ideas have inspired the field of topological photonics
\cite{haldane2008prl,ozawa_topological_2019,hafezi2023pra,activetopological2020review}, 
where propagation of light is controlled by topological effects leading to applications such as photon routing \cite{routing2018sci} and directional amplifiers \cite{peano2016prx,porras_topological_2019}. 
Photonic setups are inherently open quantum systems subjected to photon loss and gain processes, similarly to other bosonic systems like polaritons \cite{amo2009superfluidity}, and vibrational \cite{topoacustics2015prl} or phononic lattices \cite{topoacustics2015prl,kiefer2019prl}.
Topology in bosonic systems is often studied at the single-particle or semi-classical level, but less is known about the interplay between topology, interactions and dissipation.

Driven-dissipative systems \cite{Carusotto_review_2013,hartmann2016review} are ideal candidates for the study of such an interplay, and in particular, Bose-Hubbard driven-dissipative lattices  \cite{hartmann2008quantum,carusotto2008prl} have been extensively investigated. This model consists of non-linear photonic modes coupled by hopping terms and subject to both photon loss and coherent driving by an external field. 
They have been addressed by theoretical methods such as the coherent-state ansatz \cite{boite2014pra,wang_pattern_2020}; the Gutzwiller ansatz \cite{boite2013prl};  the truncated Wigner approximation \cite{vicentini_critical_2018}; Keldysh path-integral techniques
\cite{sieberer2016keldysh,grass2019excitations}; and the corner renormalization group method \cite{finazzi2015prl}, showing a non-equilibrium phase diagram with a first-order phase transition between high and low-density phases as a function of the driving strength. 
More recent works have explored critical slowing down \cite{vicentini_critical_2018}, 
entanglement \cite{Casteels2016dimer}, quantum chaos \cite{dahan_classical_2022}, and pattern formation \cite{wang_pattern_2020}.
Driven-dissipative Bose-Hubbard chains and lattices can be naturally implemented in the quantum limit of small photon occupations using superconducting circuits \cite{houck2012review, yanay2020two}, where strong-interaction effects \cite{ma2019dissipatively}, and even a dissipative quantum phase transition, 
have been experimentally observed \cite{fitzpatrick_observation_2017}. 
Other implementations are microcavity polaritons \cite{rodriguez2016interaction} and trapped ions \cite{porras2004bose,debnath2018prl}.

A natural path to induce topological phases in quantum driven-dissipative systems is to break the time-reversal symmetry. 
This approach has been recently pursued with the study of incoherently or parametrically driven linear photonic lattices \cite{porras_topological_2019, wanjura2020topological, ramos2021pra,gomez2022pra, gomez-leon_driven-dissipative_2023}, leading to the discovery of topological amplification. 
This phenomenon is intimately linked to non-Hermitian topology, since linear dissipative systems are governed by a non-Hermitian dynamical matrix \cite{nonHtopology2018prx,nonHtopology2019prx}; it is related to so called point-gap topology \cite{gomez-leon_driven-dissipative_2023, brunelli2023restoration};
and it manifests itself as a directional amplification of input signals.  

In this work we study a dissipative Bose-Hubbard chain in which time-reversal symmetry is broken by adding a site-dependent phase to the coherent drive. 
We employ a Gaussian variational ansatz to approximate the density matrix \cite{guaita2019gaussian,hackl2020scipost} and explore the emergence of non-Hermitian topological phases generated by the activation of parametric terms.
Our main results are:
(i) Stable steady-state solutions can be achieved by using an inhomogeneous profile for the coherent drive that smoothens out boundary effects.
(ii) The non-equilibrium phase diagram is divided into low and high boson density phases separated by a phase coexistence region. In the latter the system exhibits non-Hermitian topology and topological amplification properties.
(iii) Within the phase coexistence region, fluctuations build up at the frontier between topological and trivial phases. 
(iv) Correlations between sites are enhanced at the topological phase.  

\emph{Driven-dissipative Bose-Hubbard model.-}
We consider $N$ non-linear photonic modes forming a chain, and described by annihilation (creation) operators, $a_j$, ($a^\dagger_j$), whose unitary dynamics is governed by the Hamiltonian
\begin{eqnarray}
H_a &=& \sum_{j=1}^N \Delta_j a^\dagger_j a_j +
J \sum_{j=1}^{N-1}  \left( a^\dagger_{j+1} a_j e^{i \phi} + a^\dagger_{j} a_{j+1} e^{- i \phi} \right)
\nonumber \\
&+& U \sum_{j=1}^N (a^\dagger_j a_j)^2 \
+ \sum_{j=1}^N \epsilon_j \left( a_j e^{i \psi_j} + a_j^\dag e^{- i \psi_j}  \right),
\label{hamiltonian}
\end{eqnarray}
where $\Delta_j$ are local mode energies, $J$ is a uniform photonic hopping rate, $\phi$ is a constant hopping complex phase and  $U$ is a local non-linearity. 
A coherent drive acts over the array with local amplitude $\epsilon_j$ and site-dependent phase $\psi_j$.
The distinctive feature of this work is the breaking of time-reversal symmetry. We consider two possibilities: 
(i) $\phi = 0$, but constant phase gradient in the coherent drive, 
$\psi_j = j {\cal \varphi}$; or 
(ii) $\phi \neq 0$ and $\psi_j = 0$. (i) and (ii) are equivalent and related by the gauge transformation $a_j \to a_j e^{-i \psi_j}$, with $\varphi = \phi$. 
Our analysis will be carried out in gauge (ii), however, gauge (i) is more convenient for experimental implementations, since only a constant gradient of the phase of the coherent drive is required \cite{ramos2022directional}. 

The whole quantum dynamics of the chain is determined by a master equation which also includes photon decay,
\begin{equation}
{\cal L}(\rho) = - i [H_a, \rho] + 
\frac{\kappa}{2} \sum_j \left( 2 a_j \rho a_j^\dagger - a_j^\dagger a_j \rho - \rho a_j^\dagger a_j  \right).
\label{master}
\end{equation}
From here, we are mostly interested in steady-state properties ${\cal L}(\rho^\star) = 0$.

\emph{Parametric chain limit.-}
We consider time-dependent mean values of the bosonic fields $\alpha_j(t) = \langle a_j (t) \rangle$, and operators capturing the fluctuations around the mean $b_j = a_j - \alpha_j(t)$.  As a first approach to the dissipative dynamics of the model, one can assume the mean fields $\alpha_j(t)$ to solve the master equation within a coherent-state ansatz, and to converge towards a steady-state value, $\alpha_j(t \to \infty) = \alpha_j^\star$. Under this assumption, the Hamiltonian part of the dynamics of the $b_j$ fields at long times is governed by the following effective Hamiltonian (to quadratic order in the $b_j, b_j^\dagger$ operators):
\begin{eqnarray}
H_b &\approx& \sum_j \tilde{\Delta}_j b^\dagger_j b_j 
+ \sum_{j=1}^N \left( g_j  {b^\dagger_j}^2 + g_j^* {b_j}^2 \right)   
\nonumber \\
&+& J \sum_{j=1}^{N-1} 
\left( b^\dagger_{j+1}  b_j e^{ i  \phi} +
b^\dagger_{j}  b_{j+1} e^{- i  \phi} \right) 
, \label{quadratic}
\end{eqnarray}
with the linear term vanishing from the $\langle b_j \rangle = 0$ condition, and 
$\tilde{\Delta}_j = \Delta_j + 4 U |\alpha_j^\star|^2 + U$, $g_j = U (\alpha_j^\star)^2 $.

To reveal the link with non-Hermitian physics we define the set of $2 N$ fluctuation operators in the Nambu basis,
$\mathsf{b}_{\mu (= j)} = b_j$ (if $\mu = 1,\dots,N$), 
$\mathsf{b}_{\mu (= j + N)} = b^\dagger_j$
(if $\mu = N+1,\dots,2N$).
In an input-ouput formalism, the linear response of these operators is 
$\dot{\mathsf{b}}_\mu = - i \sum_\nu \mathbb{H}_{\mu \nu} \mathsf{b}_\nu + F_\mu$, with $F_\mu$ representing input fields.
Here  $\mathbb{H}_{\mu \nu} = (\mathbb{J} \mathbb{V})_{\mu \nu} - i (\kappa/2)  \delta_{\mu\nu}$ is 
the non-Hermitian matrix defined starting from Eq.~\eqref{quadratic}, 
namely $H_b = \frac{1}{2}{\mathbf{b}}^\dagger \mathbb{V} {\mathbf{b}}$, and 
$\mathbb{J}_{\mu \nu} = [b_\mu,b_\nu^\dagger]$.
We characterize the topology properties of the steady-state of our model \eqref{master} with the winding number 
\begin{equation}
\nu = \frac{1}{2\pi} {\rm Im} \int_{-\pi}^{\pi} d k \ {\rm Tr} \partial_k \log(\mathbb{H}(k)) ,
\label{winding}
\end{equation}
where $\mathbb{H}(k)$ is the Fourier transform of the non-Hermitian matrix $\mathbb{H}$, and we assume an homogeneous system. 
In \cite{gomez-leon_driven-dissipative_2023} it was shown that non-zero values of $\nu$ lead to the chain entering in a regime of directional amplification, which imply large elements of the zero-frequency Green's function, ${\mathbb{G}} = - {\mathbb{H}}^{-1} $. $\nu$ is also related to non-Hermitian point-gap topology \cite{nonHtopology2019prx} (see S\ref{Topological}, for details).
\emph{Gaussian ansatz.-}
Integrating the quantum evolution described by Eq. \eqref{master} is a challenging quantum many-body problem.
However in the limit of weak non-linearities a Gaussian ansatz is a good approximation to the exact steady-state of the system. 
Specifically, we write equations of motion for $\alpha_j$
and correlators $G_{jk} = \langle b^\dagger_j b_k \rangle$ and $F_{jk} = \langle b_j b_k \rangle$.  
The Gaussian ansatz can be applied at the level of equations of motion by using Wick's theorem on correlators of the form
$ \langle B_1 B_2 B_3 B_4 \rangle 
= 
\langle B_1 B_2 \rangle \langle B_3 B_4 \rangle 
+ 
\langle B_1 B_3 \rangle \langle B_2 B_4\rangle 
+ 
\langle B_1 B_4\rangle \langle B_2 B_3\rangle $, with $B_j = b_j$ or $b^\dagger_j$.
The ansatz has some desirable properties that have been proven in the case of closed quantum systems, namely, Gaussian states form a so-called K\"ahler manifold which implies that equations of motion can be derived consistently from variational principles \cite{hackl2020scipost,guaita2022quantum}. 
The master equation in Eq. \eqref{master} can be viewed as the system's reduced-state evolution resulting from the joint unitary dynamics of the system plus bath. Applying a Gaussian ansatz to such a joint dynamics corresponds exactly to our approximation, since the reduced state of a Gaussian state is also Gaussian. Hence, in the case of a Markovian bath, the Gaussian ansatz applied to dissipative dynamics enjoys the same properties as those for unitary dynamics (see S\ref{Gaussian} for details).

Applying Wick's theorem to the Heisenberg equations of motion, one gets
\begin{widetext}
\begin{align} 
\frac{d \alpha_j}{d t} = & 
- i \left[   \epsilon_j + (\Delta_j + U) \alpha_j + 
2 U ( |\alpha_j|^2 \alpha_j + 2 \alpha_j G_{jj} + \alpha_j^* F_{jj} )
+ J \left( \alpha_{j+1} e^{- i \phi} + \alpha_{j-1} e^{i \phi}  \right) \right] - \frac{\kappa}{2} \alpha_j , \nonumber \\ 
\frac{d G_{jk}}{dt} 
= & i U \left( 2 (\alpha_j^*)^2 F_{jk}  + 4 |\alpha_j|^2 G_{jk}   + 2 F_{jk} F_{jj}^* 
  + 4 G_{jj} G_{jk}  \right)  +  i \Delta_j G_{jk}
  + i J  (G_{j-1 k} - G_{j k+1}) e^{- i \phi}   - \frac{\kappa}{2} G_{jk}  + (j \leftrightarrow k)^* , 
  \nonumber   \\
\frac{d F_{jk}}{dt}   = & 
 - i U \left( 4 |\alpha_k|^2 F_{jk}  + 2 \alpha_j^2 G_{jk}  
 + 4 G_{jj} F_{jk} + 2 G_{kj} F_{kk}  +  F_{jk} +  \delta_{jk} (\alpha_j^2 + F_{jj})  \right)
 - i \Delta_j F_{jk}   
 \nonumber   \\ & 
 - i J \left(F_{j k+1} e^{- i \phi} + F_{j k-1} e^{ i \phi} \right) - \frac{\kappa}{2} F_{jk}
+ (j \leftrightarrow k) .
\label{eq:Gaussian}
\end{align}
\end{widetext}
In S\ref{Numerical} we benchmark the Gaussian ansatz results with exact ones for a single site. Extending the comparison to larger systems is prohibitive, due to the large size of the local Hilbert space. However, we can expect that the Gaussian ansatz is accurate as long as the field fluctuations are small compared to their averages, 
$\langle b^\dagger_j b_j \rangle / |\alpha_j|^2 \ll 1$, a condition met by our results, see S\ref{Numerical}.

\begin{figure}[t]
    \includegraphics[width=1\linewidth]{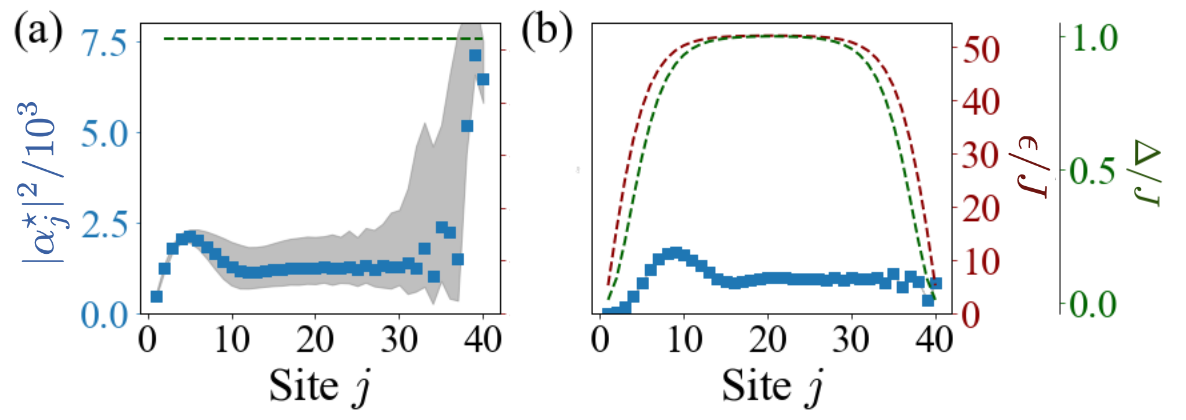}
    \caption{
    Long-time limit of $\alpha_j$ obtained with 
    (a) a homogeneous profile ($\epsilon_j = \epsilon$) and
    (b) the inhomogeneous profile in Eq. \eqref{eq.inhom}. 
    The grey area in (a) refers to the range of values that $\alpha_j$ takes in the oscillating long-time dynamics. Dashed red and green lines refer to profiles of $\epsilon_j$ and $\Delta_j$, respectively.
    $N = 40$,  $\phi =  \pi/3$, $\kappa/J = 1$, $U/J = - 2 \cdot 10^{-4}$.}
    \label{fig:profile}
\end{figure}

\emph{Stability diagram.-}
We have numerically integrated Eqs. \eqref{eq:Gaussian} with a homogeneous profile for the driving amplitude 
$|\epsilon_j| = \epsilon$ and detuning $\Delta_j = \Delta$, and found that in most of the phase diagram the system exhibits persistent fluctuations at long times, and fails to reach a steady state. 
Yet an inhomogeneous profile of the amplitude and detuning of the local drives lead to the emergence of a steady state, attenuating 
the effects of the open boundaries \cite{ramos2022directional}. Throughout the paper we choose, for $j \leq N/2$,
\begin{equation}
%\epsilon_j = \epsilon \ \tanh((j-N_0)/N_0) 
\epsilon_j = \epsilon \ \tanh(j/N_0) ~~~~~~~~  {\Delta}_j = \Delta \ \tanh^2(j/N_0)~, 
\label{eq.inhom}
\end{equation}
and a mirror-reflected pattern for $j > N/2$. $N_0$ in Eq. \eqref{eq.inhom} is the size of the border region.
Smooth boundary conditions allow us to obtain stable steady-states in a larger region of the non-equilibrium phase diagram. Fig. \ref{fig:profile} shows a typical comparison between solutions with (a) homogeneous drive (i.e. sharp boundary conditions); and (b) the smoothened boundary conditions, showing the stabilization of the steady-state (further examples in S\ref{Time}).

\begin{figure}[t]
    \centering
    \includegraphics[width=1\linewidth]{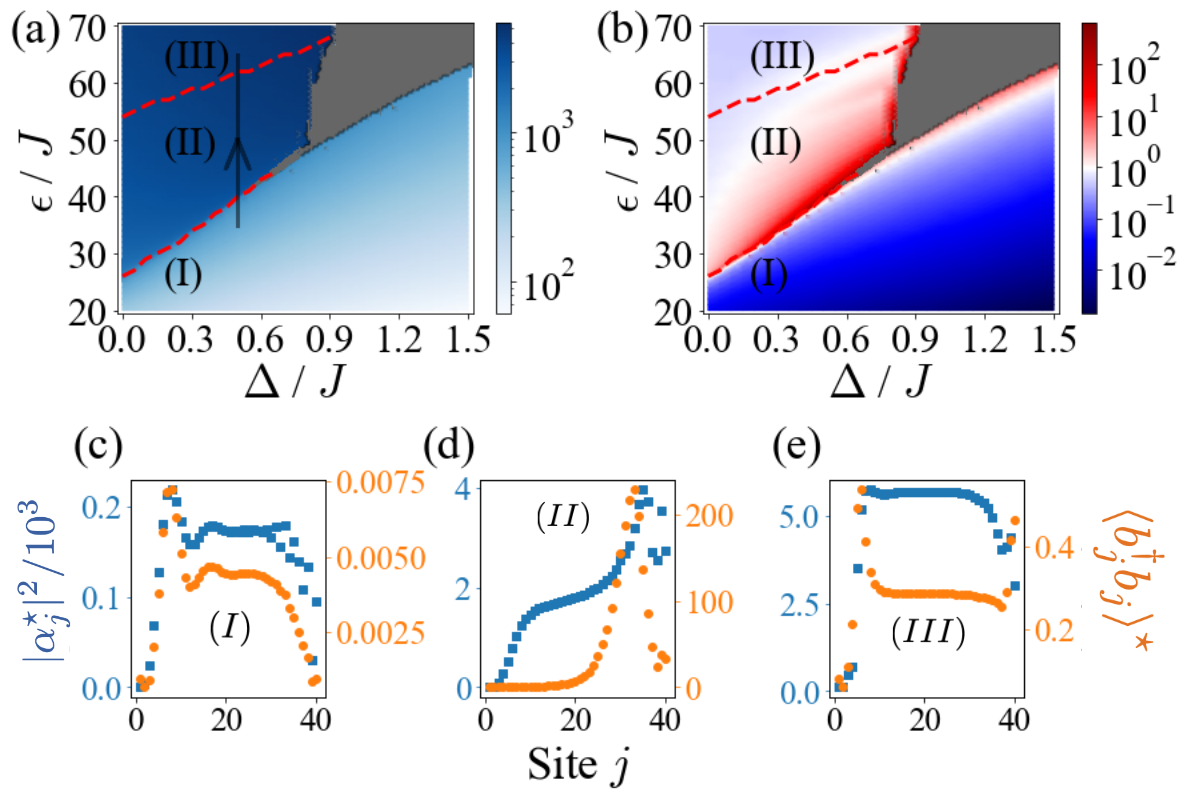}
\caption{
Phase diagram of:
(a) Mean displacement value,  $\overline{|\alpha^\star|^2} = (1/N) \sum_{j=1}^N |\alpha_j^\star|^2$, over the chain, 
(b) Maximum value of $\langle b_j^\dagger b_j \rangle^\star $ over the chain, 
(c, d, e) Steady-state profile of the chain (blue squares: $\langle b_j^\dagger b_j \rangle^\star $ and 
orange circles: $|\alpha_j^\star|^2$) for the phases 
I ($\epsilon/J = 20$),  II ($\epsilon/J = 40$) and  III ($\epsilon/J = 70$), whereas $N = 40$,  $\phi =  \pi/3$, $\kappa/J = 1$, $U/J = - 2 \cdot 10^{-4}$.
}
\label{fig:phase_diagram}
\end{figure}

\emph{Non-equilibrium steady-state phase diagram.-}
We focus on a regime of weak non-linearities 
$| U | \ll \epsilon_j$, in which the quantum dynamics is close to the effective quadratic model in Eq. \eqref{quadratic}. 
Figs. \ref{fig:phase_diagram}(a,b) show the calculated phase diagram.
In Fig. \ref{fig:phase_diagram}(a) we recognize a low density region phase (I).
The rest of the phase diagram comprises a high-density region (III) and a co-existence region (II), a division that is not apparent in Fig. \ref{fig:phase_diagram}(a), but will become clear below. This phase diagram is in sheer contrast with the $\phi = 0$, time reversal case (see S\ref{Phi}).
Finally, the grey area is a chaotic region at which the long-time behaviour of the Gaussian ansatz does not converge to a steady-state state. 
Fig. \ref{fig:phase_diagram}(b) shows that fluctuations are enhanced at the critical line separating (I) and (II).
Figs. \ref{fig:phase_diagram}(c-e) show  typical steady-states in each of the three regions, exhibiting in particular a highly inhomogeneous boson density distribution in the phase-coexistence region (Fig. \ref{fig:phase_diagram}(d)).

\begin{figure}[t]
    \centering
    \includegraphics[width=1\linewidth]{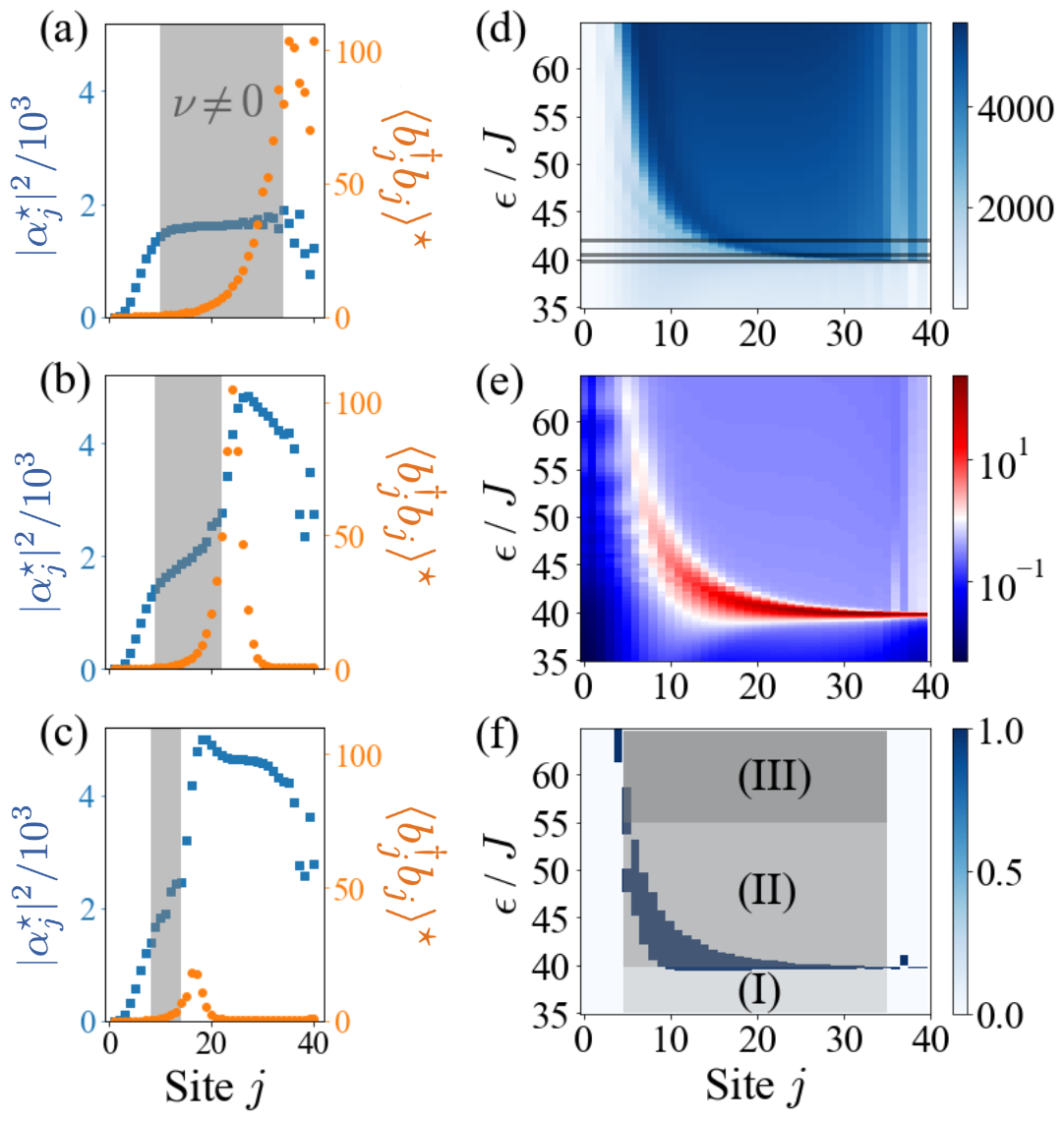}
    \caption{Steady-state profile of the chain for three values of $\epsilon$: (a) $\epsilon = 39.8/J$ , (b) $\epsilon = 40.5/J$ and (c) $\epsilon = 42/J$.
    (d) Values of $|\alpha_j^\star|^2$, (e) fluctuations $\langle b_j^\dagger b_j\rangle^\star$ and (f) local winding number $\nu_j$   when $\epsilon$ is varied along a line of constant $\Delta$ on the phase diagram in Fig. \ref{fig:phase_diagram}(a), 
    whereas $N = 40$, $\Delta = 0.5$, $\phi =  \pi/3$, $\kappa/J = 1$, $U/J = - 2 \cdot 10^{-4}$.
    }
    \label{fig:phase_transition}
\end{figure}

\emph{Topology in the phase coexistence region.-}
Let us focus on the transition between low and high-boson density phases in region (II) of the phase diagram.
The coexistence region is bounded by values 
$\epsilon_{\rm c,1}$ and $\epsilon_{\rm c,2}$ of the coherent drive as we increase $\epsilon$ along lines of constant $\Delta$.
In  Fig \ref{fig:phase_transition}(a-c) we show the evolution of density profiles as we move along the line on Fig. \ref{fig:phase_diagram}(a).
The chain is divided in two spatial regions of high and low boson densities, separated by an interface with large number of fluctuations. The interface moves from right to left as we ramp up the value of $\epsilon$.
A topological characterization can be achieved by using the definition in Eq. \eqref{winding}. 
In a homogeneous system $\mathbb{H}$ is parameterized by constant values $g_j = g$, $J$, $\Delta_j = \Delta$ and the winding number is a function $\nu(g,J,\Delta)$.
In order to characterize spatial regions in our inhomogeneous system we define the function
$\nu_j = \nu(g_j,J,\tilde{\Delta}_j)$ with the local Hamiltonian parameters. This local definition of a winding number is correct in a quasi-homogeneous limit in which parameters vary slowly in space.
Different values of $\nu_j$ across the chain imply coexistence of different topological phases, as we show below.
As we modify $\epsilon$ along a constant $\Delta$ line we find: 
(i) for $\epsilon < \epsilon_{\rm c,1}$ (region I), a low-density phase with a spatially homogeneous density profile and 
$\nu_j = 0$.
(ii) 
$\epsilon = \epsilon_{{\rm c},1}$ marks the onset of topological amplification, see Fig \ref{fig:phase_transition}a, and the transition between phases (I) and (II). 
At this critical point $\nu_j = 1$, for all $j$'s except for the edges of the chain.
(iii) 
$\epsilon_{{\rm c},2} > \epsilon > \epsilon_{{\rm c},1}$, phase coexistence region  
in which the chain is divided in a low-density, topological spatial region with $\nu_j = 1$ that is continuously 
shrunk and replaced by a high-density topologically trivial region with $\nu_j =0$, 
see Figs \ref{fig:phase_transition}(b,c). 
The topological and trivial spatial regions are separated by a moving edge with a large density of fluctuations. 
The continuous evolution and phase coexistence is clearly seen in Fig \ref{fig:phase_transition}(d), 
as well as the fluctuation maxima at the interface between topological and trivial phases in 
Fig \ref{fig:phase_transition}(e). Fig. \ref{fig:phase_transition}(f) shows the evolution of the topological region as identified by $\nu_j$.
(iv) Finally for values 
$\epsilon > \epsilon_{{\rm c},2}$ (region III) the whole chain enters into the high-density, non-topological phase.
A quantitative definition of the boundary between (II) and (III), $\epsilon_{{\rm c},2}$, is complicated by finite-size effects. 
In order to establish a criterion we define $\epsilon_{{\rm c},2}$ as the point at which $\nu_j$ vanishes at all points $j$ within the bulk of the chain, corresponding to the segment $N_0 < j < N - N_0$.

\begin{figure}[t]
    \centering
    \includegraphics[width=1\linewidth]{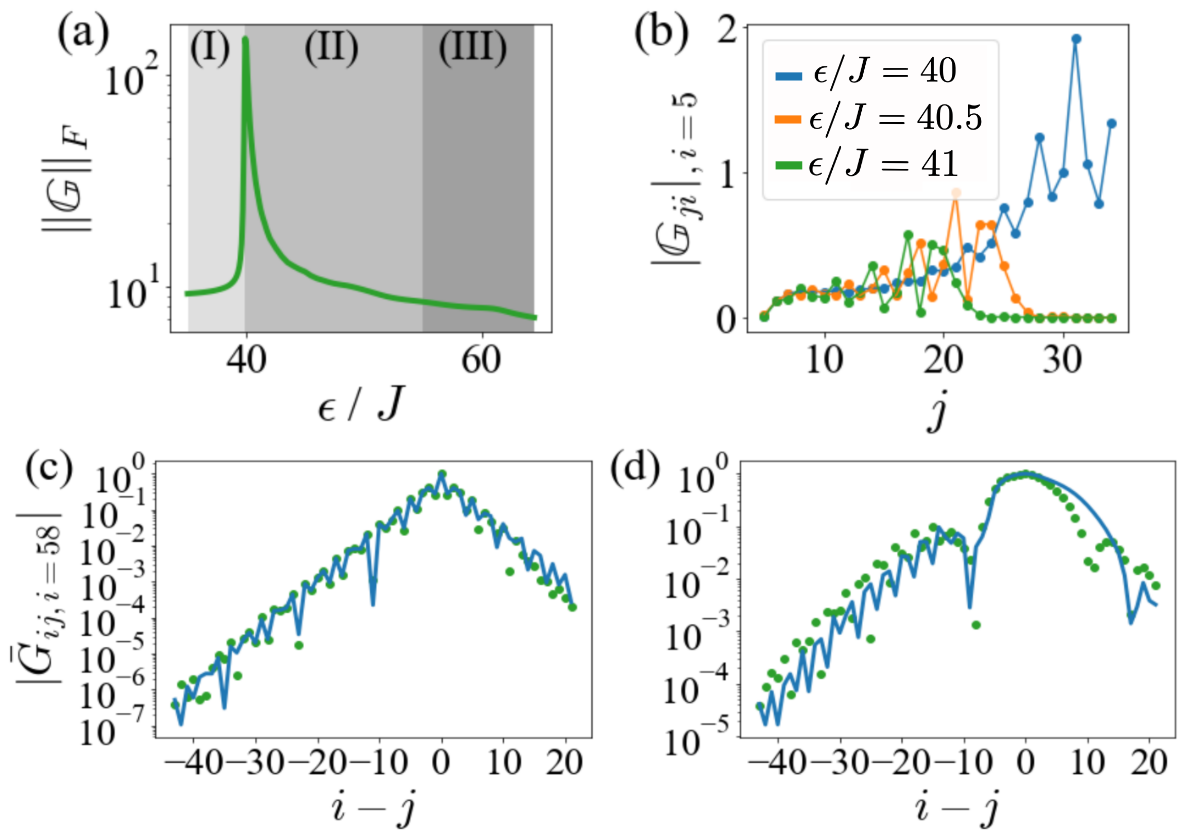}
    \caption{(a) Norm of the Green's function 
    $||\mathbb{G}||_F = 
    \sqrt{{\rm Tr}(\mathbb{G}^\dagger \mathbb{G})}$ and 
    (b) values of some non-diagonal elements of the green function for $N=40$, $\Delta/J = 0.5$, $\phi = \pi/3$, $\kappa/J = 1$ and $U/J= -2 \cdot 10^{-4}$.
    Normalized correlation function $\bar{G}_{jk}$ (c) $N = 80, \epsilon/J = 38$ (low density phase), (d) $N = 80, \epsilon/J = 40.2$ (topological coexistence phase) for the non-linear Bose-Hubbard chain (green dots) and the effective quadratic model extract from the steady-state value (blue line). $\Delta/J = 0.5$, $\phi = \pi/3$, $\kappa/J = 1$ and $U/J= -2 \cdot 10^{-4}$.
    }
    \label{fig:correlations}
\end{figure}

\emph{Green's function and correlation functions.-}
Our calculations show that both the chain Green's function $\mathbb{G}$ and the correlations are enhanced at the topological phase. Topological properties of $\mathbb{G}$ can be derived by the theory of topological amplification (see S\ref{Topological}). 
This is indeed what we observe in Fig. \ref{fig:correlations}(a). 
Fig. \ref{fig:correlations}(b) also shows that, in the topological region of the chain, 
for a site $i$ on the left-hand side of the chain, $\mathbb{G}_{ji}$ grows with $j$, which is a signature of directional amplification.
We have also studied normalized two-point correlation functions 
$\bar{G}_{jk} = G_{jk}/\sqrt{n_j n_k}$ at the low density phase and at the critical line separating phases (I) and (II) for a chain of $N = 80$ sites.
In Figs. \ref{fig:correlations} (c,d) we observe longer-range correlations on the critical line, whereas at the low density phase, correlations decay exponentially with distance.

\emph{Conclusions and Outlook.-} 
We have shown that a driven-dissipative Bose-Hubbard model with a site-dependent, linearly varying phase in the coherent drive exhibits topological amplification at the phase boundary between a high- and low-density phase. Topological phases can be characterized by a topological invariant associated to point-gap topology. Our model shows the emergence of non-Hermitian topology and its physical consequences in a quantum \emph{non-linear} dissipative system. 

Our model can be implemented with arrays of superconducting cavities, following the scheme in Ref.~\cite{ramos2022directional}, in which the Kerr non-linearity $U$ arises naturally from Josephson junctions and dissipation $\kappa$ from the coupling to external transmission lines. In addition, the phase gradient $\phi$ and the inhomogeneous profiles of amplitude $\epsilon_j$ and detunings $\Delta_j$ are induced via the external microwave pump and a local tuning of the cavity frequencies. Our work may lead to the design of non-linear amplifiers and photo-detectors in the microwave regime. Future work will involve the characterization of critical properties and time scales, the study of non-Gaussian regimes, and extensions to two dimensions where topological amplification can occur in chiral edge-states \cite{vega2024topological}.

\emph{Acknowledgements.-}
DP and T. Ramos acknowledge support from 
Spanish projects PID2021-127968NB-I00 funded by MICIU/AEI/10.13039/501100011033, 
by Proyecto Sin\'ergico CAM 2020 Y2020/TCS-6545 (NanoQuCo-CM), and the CSIC Research Platform on Quantum Technologies PTI-001. T. Ramos further acknowledges support from the Ramón y Cajal program RYC2021-032473-I, financed by MCIN/AEI/10.13039/501100011033 and the European Union NextGenerationEU/PRTR. LR and T. Roscilde acknowledge the support of PEPR-q (QubitAF project).

%%%%%%%%%%%%%%%%%%%%%%%%%%%%%%%%%%
\bibliography{references}
\bibliographystyle{apsrev4-2}

\clearpage
\vspace{0.5cm}

\onecolumngrid
\begin{center}
  \vspace{1cm}
  {\large\textbf{SUPPLEMENTARY MATERIAL}}
 \vspace{1cm}
\end{center}

\twocolumngrid
%\begin{center}
%Pablo D\'iez-Valle$^1$, Fernando Mart\'inez-Garc\'ia$^1$, Juan Jos\'e Garc\'ia-Ripoll$^1$, and Diego Porras$^1$

%\textit{$^1$Instituto de F\'{\i}sica Fundamental IFF-CSIC, Calle Serrano 113b, Madrid 28006, Spain}\\
%\textit{$^1$Instituto de F\'{\i}sica Fundamental IFF-CSIC, Calle Serrano 113b, Madrid 28006, Spain}
%\end{center}

\section{Gaussian ansatz with dissipation}
\label{Gaussian}
In the case of closed quantum systems, Gaussian wavefunctions form a K\"ahler manifold, which implies
that the equations of motion obtained by applying Wick's theorem can be consistently derived from variational principles (see \cite{guaita2019gaussian,guaita2022quantum}).
Here, we show that the same variational principles justify the Gaussian ansatz in our dissipative case. 
Our argumentation relies on the fact that Eq. (\ref{master}) can be obtained by replacing dissipation 
by the quadratic coupling to a Gaussian bath.  
The Gaussian ansatz can be applied to the joint system-bath unitary dynamics. Since the reduced
state of a Gaussian state is also Gaussian state, this means that the dissipative dynamics described by Eq. (\ref{master}) can also be well approximated by a Gaussian ansatz. 
We confirm this argumentation by showing that applying the Wick's theorem to the system-bath equations of motion is equivalent to Eqs. \eqref{eq:Gaussian}, within the Markovian approximation.

We consider unitary dynamics in an enlarged Hilbert space by defining a set of $M$ bath bosonic operators 
$c_{j n}$, $c^\dagger_{j n}$, $n = 1, \dots, M$, for each site $j$, such that the Hamiltonian becomes
\begin{equation}
H = H_a + H_{\rm I} + H_{\rm B},
\end{equation}
where $H_a$ is the bosonic Hamiltonian in Eq. (\ref{hamiltonian}) and 
\begin{eqnarray} 
H_{\rm B} &=& \sum_{j,n} \omega_n  c^\dagger_{n j} c_{n j} , \nonumber \\
H_{\rm I} &=& \sum_{j,n} g_{n} \left( a_j^\dagger c_{n j} + a_j c_{n j}^\dagger \right) .
\end{eqnarray}
We assume that at time $t = 0$ the bath is in the vacuum $c_{j n} |\Omega\rangle_{\rm c} = 0$. 
For simplicity we consider that all local baths have the same coupling function $g_n$ and linear dispersion $\omega_n = v_c n$, with $n = -M/2, -M/2 + 1, \dots, M/2$.
The latter assumes a bath with negative energies, which can be understood as originating from a change of rotating frame and a redefinition of the zero of energies in the bath's modes.
In any case, the thermodynamical stability of the bath is not required in the following discussion, which solely relies on the equivalence between the non-equilibrium dynamics under $H$ and the master equation.

Consider that the total system starts off in the initial state
\begin{equation}
|\Psi(0) \rangle = |\psi_a\rangle |\Omega\rangle_{\rm c} .
\label{eq:initial}
\end{equation}
Our aim is to write the equations of motion including the bath and for this it will be advantageous to work in the interaction picture with respect to $H_{\rm B}$,
\begin{equation}
H_{\rm I}(t) = \sum_{j,n} g_n \left( a_j^\dagger c_{jn} e^{-i\omega_n t} + a_j c^\dagger_{jn} e^{-i\omega_n t}  \right).
\end{equation}
Now we can write down the equations of motion,
\begin{eqnarray}
    \dot{\alpha}_j &=& - i \sum_n g_n \langle c_{nj} \rangle e^{-i \omega_n t} +  \dot{\alpha}_j|_{H_a} \nonumber, 
\\
    \dot{c}_{nj} &=& - i \sum_n g_n a_j e^{i \omega_n t} ,  \nonumber \\
    \dot{G}_{jk} &=&  i \sum_n g_{jn} e^{i \omega_n t} \langle c^\dagger_{nj} b_k (t) \rangle 
    - i g_{jn} e^{-i \omega_n t} \langle b_j^\dagger c_{nk} \rangle \nonumber  \\ &+&  \dot{G}_{jk}|_{H_a}  ,
    \label{eq:evc}
\end{eqnarray}
where $\dot{A}|_{H_a}$ represents evolution under $H_a$. 
We use Eq. \eqref{eq:evc} together with the condition in Eq. \eqref{eq:initial} such that we get
\begin{eqnarray}
    \dot{\alpha}_j 
    &=& - \sum_n g^2_n \int dt' \langle c_{nj} (t') \rangle e^{-i \omega_n (t-t')} +  \dot{\alpha}_j|_{H_a} , \nonumber \\
    \dot{G}_{jk} &=&  - \sum_n g^2_{n} \int_0^t e^{i \omega_n (t-t')}  \langle a^\dagger_j(t') b_k(t) \rangle dt'  \\ 
                 & &  - \sum_n g^2_{n} \int_0^t e^{-i \omega_n (t-t')} \langle b_j^\dagger(t) a_{k}(t') \rangle dt' +  \dot{G}_{jk}|_{H_a} \nonumber ,
\end{eqnarray}
with a similar expression for $F_{jk}$. 
So far, the assumption of a Gaussian ansatz only enters into the calculation of $\dot{G}_{jk}|_{H_a}$, and $\dot{\alpha}_j|_{H_a}$. 
Since we are considering a unitary system plus bath evolution, the Gaussian ansatz and the application of Wick's theorem to the above set of equations can be derived variationaly.
To finally arrive at our Eq. (\ref{eq:Gaussian}) we assume the approximation \cite{breuer2002theory}
\begin{equation}
\sum_n g^2_{n} e^{- i \omega_n (t-t')} = \frac{\kappa}{2} \delta(t-t') ,
\end{equation}
which holds within the usual Born-Markov limit, $\kappa \tau_m \ll 1$ with $\tau_m$ the memory time of the function
$\sum_n g^2_{n} e^{- i \omega_n (t-t')}$.
There is thus a formal equivalence between the equations of motion of the dissipative system and those equations derived form the closed unitary dynamics with a Gaussian variational ansatz. 

\section{Numerical study of the validity of the Gaussian ansatz}
\label{Numerical}
\begin{figure}[t]
    \centering
    \includegraphics[width=1\linewidth]{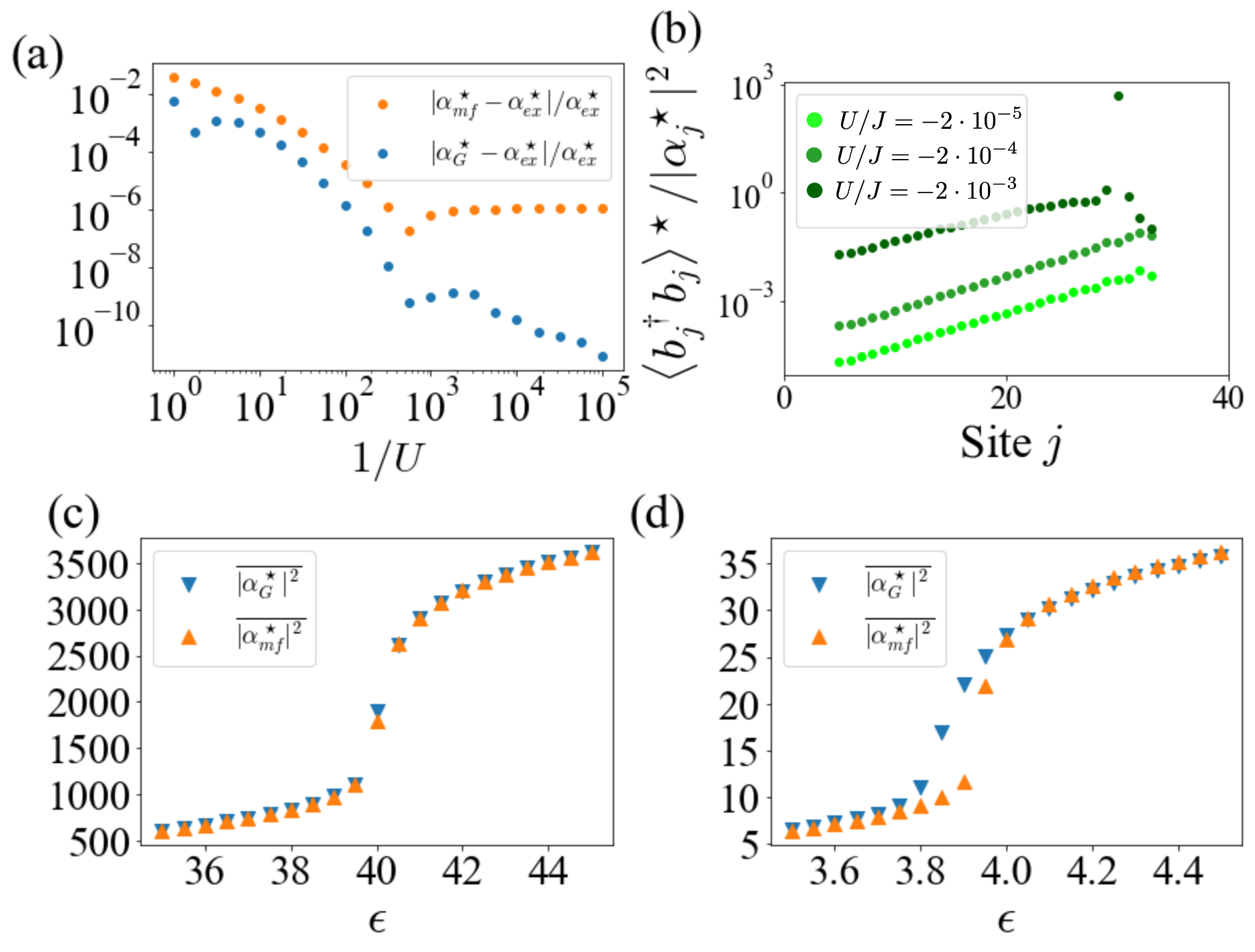}
    \caption{
    (a) Relative error for $\alpha^\star = \langle a \rangle (t \to \infty)$. Blue dots:  
    $|\alpha_{\rm G}^\star - \alpha_{\rm ex}^\star |/ |\alpha_{\rm ex}^\star|$, with $\alpha_{\rm G}^\star$ the Gaussian ansatz result and 
    $\alpha_{\rm ex}^\star$ the exact result. 
    Orange dots: $|\alpha_{\rm mf}^\star - \alpha_{\rm ex}^\star|/|\alpha_{\rm ex}^\star|$, 
    with $\alpha_{\rm mf}^\star$ the mean-field result, and $\epsilon, \Delta, \kappa = 1$.
    (b) Maximum value of $\langle b_j^\dagger b_j \rangle^\star  / |\alpha_j^\star|^2 $ for each sites through the phase transition, at $\Delta = 0.5$, and $U = -0.02$, $-2 \cdot 10^{-4}$, $-2\cdot 10^{-5}$ from dark to light green (units such that $J = 1$), whereas $N=40$, $\phi = \pi/3$, $\kappa = 1$. 
    (c, d) Comparison between the average of the steady-state displacement along chains, 
    $\overline{|\alpha^\star|^2} = (1/N) \sum_{j=1}^N |\alpha_j^\star|^2$ calculated with the full Gaussian ansatz (blue) and a mean-field approximation (orange) for (c) $U=-2 \times 10^{-4} $ and (d)  $U=-2 \times 10^{-2} $, whereas $N=40$, $\Delta = 0.5$, $\phi = \pi/3$, $\kappa = 1$. }
    \label{fig:SM1}
\end{figure}

We check the validity of our Gaussian ansatz by: (i) comparing its results with exact numerical diagonalizations and (ii) calculating the ratio between fluctuations and coherent bosons.
First, we benchmark the Gaussian method described in the main text with an exact diagonalization for a single site problem. 
Since we must deal with very large boson occupation numbers, exact numerical calculations become prohibitive already for $N \geq 2$ chains.
In Fig \ref{fig:SM1}(a) we plot the relative error of the mean value of the steady-state coherence,
$| \alpha^\star |$, as $U$ is varied for $N=1$ cavity. 
We plot both a mean field approach and the Gaussian ansatz method, and numerically confirm that the Gaussian ansatz is very accurate within the range of interaction strengths used in this work, with the Gaussian ansatz implying a correction over the results obtained by the mean-field ansatz.

In the  case of $N$ cavity chains, the problem is too complex to obtain the exact solution numerically. 
Still, we can find numerical evidence that the Gaussian approximation is still valid by estimating the relative strength of non-linear terms, which can be done by calculating the ratio between 
Gaussian fluctuations $\langle b_j^\dagger b_j \rangle^\star$ and number of coherent bosons in the steady-state 
$|\alpha_j^\star|^2$. 
This comparison is presented in Fig \ref{fig:SM1}(b), where we have chosen values at the phase transition between phases (I) and (II), something that represents the worst-case scenario.
The Gaussian approximation is well justified for the values of $U$ that we have chosen in this work.

Finally, \ref{fig:SM1}(c,d) shows a comparison between the Gaussian ansatz prediction and a simple mean-field calculation that neglects correlations. The two calculations agree for small $U$ (panel (c)) but depart from each other close to the phase transition at large values of $U$ (panel (d)), as expected.
\section{Time-dependent numerical results}
\label{Time}
So far we have focused on steady-state results. 
However, our calculations show that the system not always reaches a steady-state in the long-time limit. In this section, we present showcases of the system's temporal evolution to illustrate this behavior and to show typical time-scales.

We study the non-equilibrium steady-state by numerically integrating the system of coupled non-linear equations in Eq. (\ref{eq:Gaussian}), always choosing photon vacuum as an initial state. 
It is actually possible that different steady-states are obtained with different initial conditions, since nonlinear equations such as Eq. (\ref{eq:Gaussian}) can have multiple long-time solutions \cite{delpino2024limit}. 
The initial photon vacuum is however well-motivated by typical experimental conditions. Depending on the system parameters, we find a stable steady-state, an oscillatory solution or an unstable long-time limit. 

\begin{figure}[t]
    \centering
    \includegraphics[width=1\linewidth]{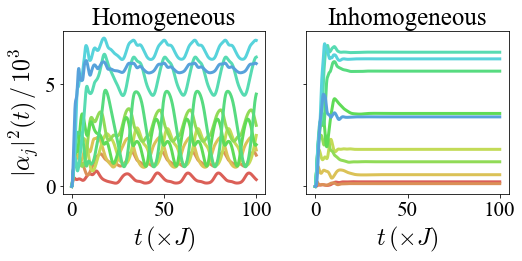}
    \caption{Dynamics of a chain of $N = 10$ cavities with a homogeneous profile of drive and cavity frequencies (left) and with an inhomogeneous profile (right). $\epsilon/J = 50$ , $\Delta/J = 0.75$ , $\phi =  \pi/3$,  $\kappa/J = 1$, $U/J = -2 \cdot 10^{-4}$.
    }
    \label{fig:SM2}
\end{figure}

In Fig. \ref{fig:SM2} we show two examples with small ($N = 10$ chains) of such time-dependent evolution to illustrate the situations that we typically find when computing the system's phase diagram.
Fig. \ref{fig:SM2} (a) presents results with a homogeneous coherent drive, in which a steady-state limit is not reached.
We attribute this behavior to the effect of sharp boundary conditions that enhance finite-size effects, which become particularly relevant in the regime of topological amplification.
Fig. \ref{fig:SM2} (b) shows time-evolution with the same set of parameters but with an inhomogeneous profile 
following Eq. (\ref{eq.inhom}), leading to a stable, not-oscillating steady-state.
It can be observed that a steady state is reached at typical values $t \approx 10/J$. 

\begin{figure}[t]
    \centering
    \includegraphics[width=1\linewidth]{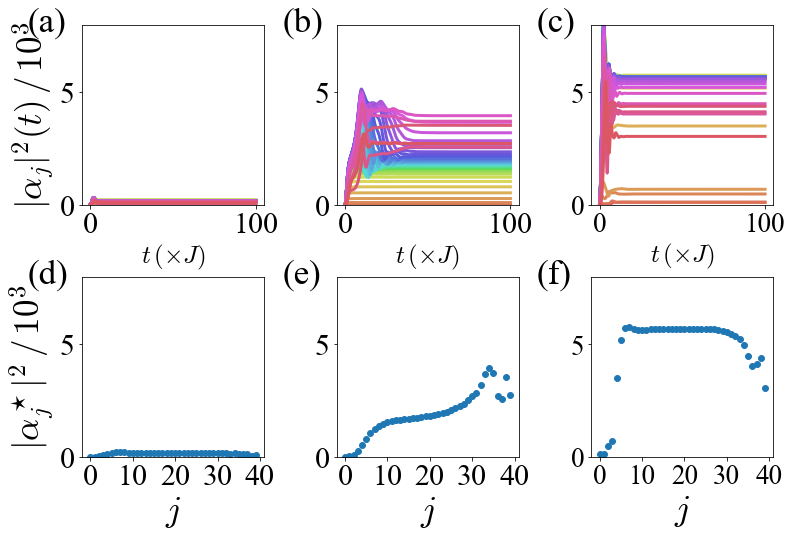}
    \caption{Dynamics of a chain of $N$ = 40 cavities (top) and their steady state (bottom) for (a,d) the low density phase $\epsilon/J = 20$ (I), (b,e) the coexistence phase $\epsilon/J = 40$ (II), and (c,f) the high density phase $\epsilon/J = 70$ (III). For all these simulations, $\Delta/J = 0.5$ ,  $\phi =  \pi / 3$ , $\kappa/J = 1$ , $U/J = -2 \cdot 10^{-4}$.
    }
    \label{fig:SM2_bis}
\end{figure}

In Fig \ref{fig:SM2_bis} we show a few examples of the dynamics of an $N = 40$ sites chain with an inhomogeneous profile in different regions of the phase diagram: (a) in the low density phase (I), (b) in the coexistence phase (II) and (c) in the high density phase (III). The stable steady-state at long time of the dynamics is shown in Fig \ref{fig:SM2_bis}(d, e, f). We see that there is numerical evidence for the slowing down of the dynamics in Fig. \ref{fig:SM2_bis} (b), with a steady-state  being generated at times $t J \approx 30$. 

%In the unstable phase, we show that the system is unstable, where both the fluctuation and the mean displacement diverges due to too large parametric amplification. In this regime, the Gaussian Ansatz is no longer a good approximation to the state of the system, as the quantum fluctuations should lead to saturation effects of the amplification at some point.

\section{$\phi = 0$ Phase diagram}
\label{Phi}

We present a brief description of the physics of the chain of cavities in the $\phi = 0$ case, for comparison with the far less trivial $\phi \neq 0$ situation.
We show that the system undergoes an abrupt phase transition from a low density phase to a high density phase when we increase the drive amplitude $\epsilon $ and at the same critical value for all sites.
There is no longer any coexistence phase, neither topological amplification phase, and fluctuations remain small all along the phase diagram.

\begin{figure}[h]
    \centering
    \includegraphics[width=1\linewidth]{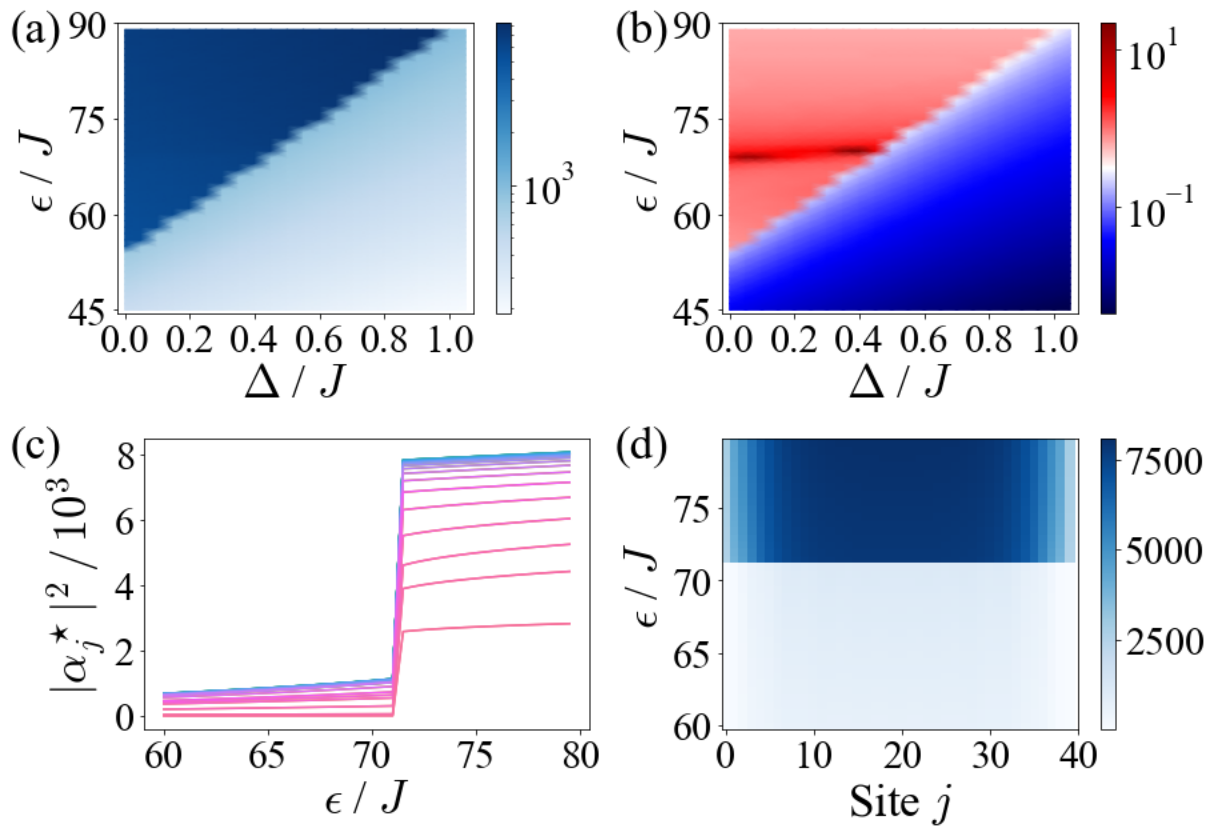}
    \caption{Phase diagram of the chain of $N=40$ sites for $\phi = 0$: (a) Mean displacement $\overline{|\alpha^\star|^2}$ (b) Maximum of fluctuations $\langle b_j^\dagger b_j \rangle^\star$
    (c),(d) Steady-state value of  $|\alpha_j^\star|^2$ for each sites through the phase transition when $\Delta/J = 0.5$ is fixed, and $\kappa/J = 1$, $U/J = -2 \cdot 10^{-4}$. 
    }
    \label{fig:SM_phi}
\end{figure}

Fig \ref{fig:SM_phi}(a,b) show the phase diagram of an $N = 40$ sites chain with $\phi = 0$. 
For a fair comparison we are using the inhomogeneous profile for the driving amplitude and detuning in 
Eq. (\ref{eq.inhom}). However, we note that a calculation with a homogeneous profile would yield the same results since,
not being subjected to the phenomenon of topological amplification, the chain achieves a long-time steady-state more easily. 
We observe in Fig. \ref{fig:SM_phi}(b) that boson fluctuations are much smaller 
(always in the range $\langle b^\dagger_j b_j \rangle^\star \leq 20$), in clear contrast with the $\phi \neq 0$ result in 
Fig. \ref{fig:phase_diagram} and \ref{fig:phase_transition}, which show much larger values. 
The phase transition for fixed $\Delta$ is illustrated both in Fig \ref{fig:SM_phi}(c) with a line of phase transition for each site of the chain and in Fig \ref{fig:SM_phi}(d) for the full chain. 
The difference between each site is mainly due to the inhomogeneous profile of the drive amplitude.

\section{Phase coexistence region and finite-size results}
\label{Phase}
We now present more detailed numerical results to shed light onto the properties of the phase co-existence region. 
In particular, we dissect the evolution of the system at each site as we cross the phase coexistence region (II).

\begin{figure}[t]
    \centering
    \includegraphics[width=1\linewidth]{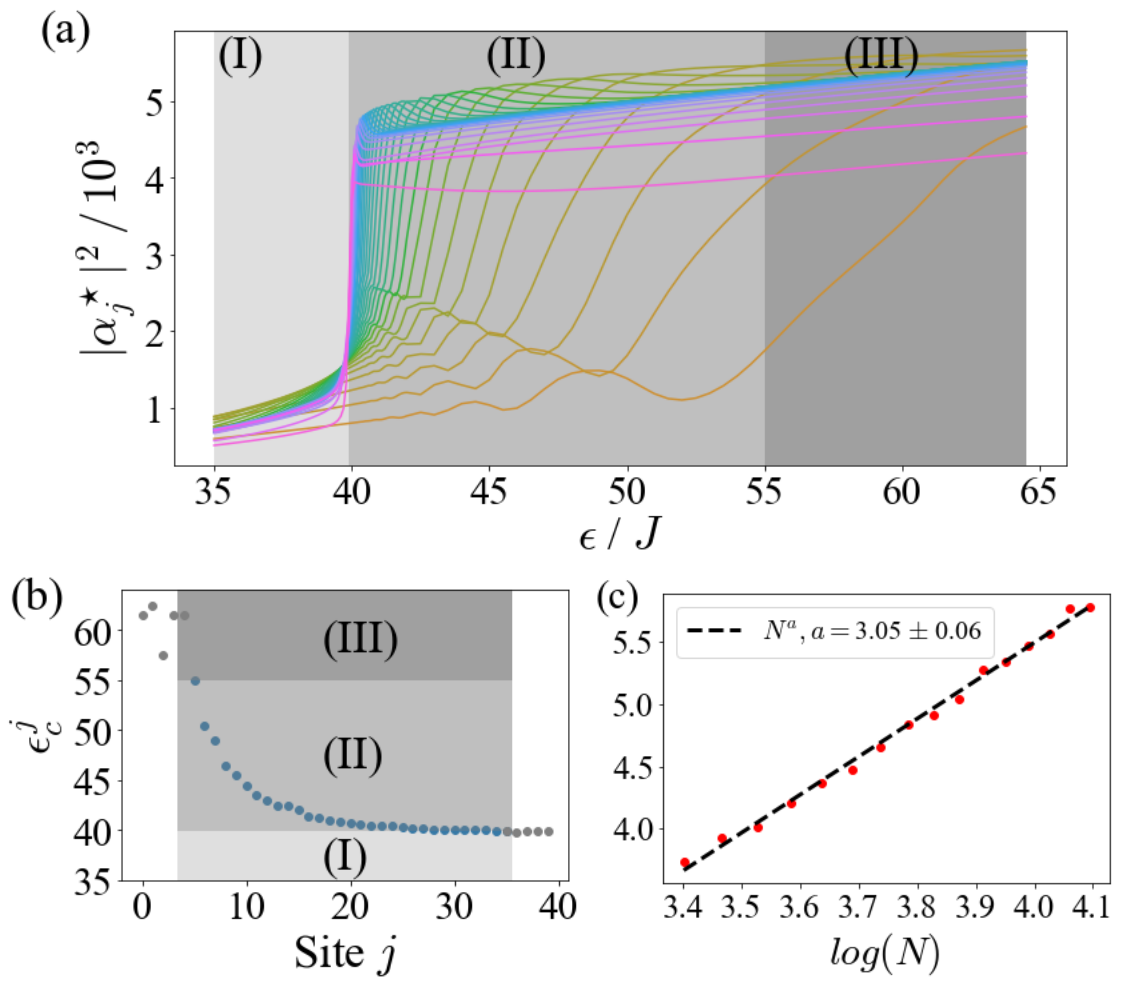}
    \caption{(a) Steady-state value of $|\alpha_j^\star|^2$ for each site of the chain through the phase transition between the low (I) and high (III) density phase.
    (b) Critical value $\epsilon_c^j$ for each site of the chain, defines as the drive that maximize the gradient of $|\alpha_j^\star|$ through the phase transition.
    (c) Finite size scaling of the derivative of $| \alpha_{N/2}^{\star} |$ as a function of $\epsilon$ at the critical point, for $N$ between $30$ and $60$ sites. $\Delta/J = 0.5$, $\phi = \pi/3$, $\kappa/J = 1$, $U = -2 \cdot 10^{-4}$ in all calculations.}
    \label{fig:SM3}
\end{figure}

In Fig. \ref{fig:SM3}(a) we show the local evolution of the steady-state values of the amplitude of the boson coherence, $|\alpha^\star_j|$. 
Locally, the system undergoes a sharp transition at a value of $\epsilon_{{\rm c}_1}^j$ that depends on the position along the chain, as discussed in the main text.
We defined $\epsilon_{{\rm c}_1}^j$ in a more quantitative manner as the value at which 
the gradient of $|\alpha^\star_j|$ as a function of $\epsilon$ is maximized. 
This value is shown in Fig \ref{fig:SM3}(b), where it is evident that, under such definition, $\epsilon_{{\rm c}_1}^j$
takes larger values for smaller values of $j$, thus confirming our observation that the frontier between the low and high density phases sweeps from right to left as we move from phase (I) into phase (III).
The analysis of the phase co-existence region is complicated by our choice of coherent drive spatial profile. Bulk properties are thus only truly defined for points in the bulk $N_0 < j < N - N_0$, as can be seen in Fig. \ref{fig:SM3}(b), where the grey points represent points at the border where the regular behaviour of the critical point with the position is lost.
We finally address the local characterization of the driven-dissipative phase transition and investigate the behaviour 
as a function of $N$ at large values of $N$. Our aim is to confirm the existence of a first-order phase transition, which we will identify as a point at which the first derivative of $\alpha_j^{\star}$ as a function of $\epsilon$ diverges in the thermodynamic $N \to \infty$ limit. 
We focus on the  middle of the chain, $|\alpha_{N/2}^{\star}|$, and calculate the derivative 
$d|\alpha_{N/2}^{\star}|/d\epsilon|_{\epsilon = \epsilon_{{\rm c}_1}^{j = N/2}}$, that is, we compute the derivate at values of $\epsilon$ at which the site $j = N/2$ undergoes the phase transition between low and high density phases. The result is plotted in Fig. \ref{fig:SM3}(c), which shows that 
\begin{equation}
\left. \frac{d|\alpha_{N/2}^{\star}|}{d\epsilon} \right|_{\epsilon = \epsilon_{{\rm c}_1}^{j = N/2}} \propto N^a
\end{equation}
with $a = 3.05 \pm 0.06$. Such powerlaw dependence is a signature of critical behaviour and confirms that the first derivative diverges in the thermodynamic limit.

\section{Topological amplification phases}
\label{Topological}
We review here the concept of topological amplification in Gaussian systems. This discussion is a summary of our previous works in \cite{porras_topological_2019,ramos2021pra,gomez-leon_driven-dissipative_2023},

The starting point is the dynamical matrix $\mathbb{H}$ that governs the input-output relations within the quadratic approximation
in Eq. \ref{quadratic}, 
\begin{equation}\label{hnh}
    \mathbb{H} = 
    \begin{pmatrix} 
    D  - i \frac{\kappa}{2} \mathds{1} &  K 
    \\
    -K^* &  -D^*  - i \frac{\kappa}{2} \mathds{1},
    \end{pmatrix},
\end{equation}
with
\begin{eqnarray}
D_{jk} &=& \Delta_j \delta_{jk} + J ( \delta_{j,k-1} e^{i \phi} + e^{-i \phi}  \delta_{j,k+1})
\nonumber \\
K_{jk} &=& 2 g_j \delta_{jk}
\end{eqnarray}
$\mathbb{H}$ determines many steady-state properties through the Green's function, 
\begin{equation}
\mathbb{G}(\omega) = \frac{1}{\omega \mathds{1} - \mathbb{H}}.
\end{equation}
The Green's function governs the linear response function of the system when subjected to perturbations at frequency $\omega$. In particular, an input field constant in time with amplitude $\bar{\epsilon}_j$ induces coherences 
$\alpha_i = \sum_j \mathbb{G}(0)_{ij}  \bar{\epsilon}_j $ in the steady-state.

A convenient topological analysis of the properties of the system can be carried out by means of the singular value decomposition (SVD), $\mathbb{H} = U S V^\dagger$, 
where $U$, $V$ are unitary matrices and $S$ is a diagonal matrix with non-negative elements, $s_n$ The Green's function at $\omega = 0$ can be written in terms of the SVD of $\mathbb{H}$,
\begin{eqnarray}
\mathbb{G}(0) = - V S^{-1} U^\dagger .
\label{sm:svd}
\end{eqnarray}
A remarkable consequence of Eq. \eqref{sm:svd} is that small singular values can dominate the Green's function, since the latter depends on $s_n^{-1}$. 
This is very important since quasi-zero singular values can appear in the SVD of the dynamical matrix, $\mathbb{H}$.  
Such quasi-zero singular values are topologically protected and they appear as a consequence of the equivalence between the SVD of ${\mathbb H}$ and the eigenvalues of an extended Hermitian matrix, 
\begin{equation}
{\cal H} = 
    \begin{pmatrix} 
    0 &  \mathbb{H} 
    \\
    \mathbb{H}^\dagger & 0
    \end{pmatrix} .
\label{eff.H}
\end{equation}
As shown for example in \cite{porras_topological_2019}, eigenvalues of ${\cal H}$, come in pairs
$E_n = \pm s_n$, where $s_n$ are the singular values of $\mathbb{H}$,
since ${\cal H}$ has a chiral symmetry, 
${\cal S}^{-1} {\cal H} {\cal S} = -{\cal H}$, with
\begin{equation}
{\cal S} = 
    \begin{pmatrix} 
    \mathds{1} &  0 
    \\
    0 & - \mathds{1}
    \end{pmatrix} .
\end{equation}
This chiral symmetry protects topological zero-energy edge states for certain symmetry classes, which typically appear if 
${\cal H}$ in Eq. $\eqref{eff.H}$ breaks time-reversal symmetry. 
Following the bulk-boundary correspondence  non-zero values of the topological invariant in Eq. \eqref{winding} imply the existence of such zero-energy states, and, thus,  the existence of singular values close to zero in expression \eqref{sm:svd}, leading to exponential amplification of the Green's function. 

To describe such topological amplification more quantitatively, let us consider the Frobenius norm of the Green's function, which can be written like

\begin{eqnarray}
\| \mathbb{G}(0) \|_{\rm F} = \sqrt{{\rm Tr} \left( \mathbb{G}(0)^\dagger \mathbb{G}(0) \right) }
= \sqrt{\sum_n \frac{1}{s_n^2}}.
\label{sm:GTrace}
\end{eqnarray}
Values $\nu = 1$ lead to the existence of a pair of zero-energy modes of $\cal H$, which translate into a zero singular-value, $s_0 \propto e^{-N/\xi}$, separated by the rest of eigenvalues by a singular value gap 
(see \cite{porras_topological_2019, ramos2021pra, gomez-leon_driven-dissipative_2023}).
In a homogeneous system with open boundary conditions the length $\xi$ is the spatial extent of the zero-energy edge-state and $N$ is the length of the system. In such non-trivial topological regime the Green's function, and thus the response of the system is exponentially enhanced, since 
\begin{eqnarray}
\| \mathbb{G}(0) \|_{\rm F} \approx \frac{1}{s_0} \propto e^{N/\xi}.
\label{sm:G0}
\end{eqnarray}
In our chiral driven dissipative Bose-Hubbard chain, we expect that the topological amplification effect will appear within a spatial region such that the local effective Hamiltonian parameters have an associated winding number $\nu \neq 0$. 

%%%%%%%%%%%%%%%%%%%%%%%%%%%%%%%%%%

\end{document}